\documentclass[12pt,preprint]{aastex}

\shorttitle{ RS Ophiuchi at radio frequencies $< 1.4$ GHz}
\shortauthors{Kantharia et al.}

\begin{document}

\title{GMRT Observations of the 2006 outburst of the Nova RS Ophiuchi: 
First detection of emission at radio frequencies $< 1.4$ GHz}

\author{N. G. Kantharia}
\affil{National Centre for Radio Astrophysics, Tata Institute of Fundamental
Research, Post Bag 3, Ganeshkhind, Pune 411 007, India}
\email{ngk@ncra.tifr.res.in}

\author{G. C. Anupama, T. P. Prabhu and S. Ramya}
\affil{Indian Institute of Astrophysics,  Koramangala, Bangalore 560 034, India}
\email{gca@iiap.res.in, tpp@iiap.res.in, ramya@iiap.res.in}

\author{M. F. Bode}
\affil{Astrophysics Research Institute, Liverpool John Moores University, Birkenhead CH41 1LD,
UK}
\email{mfb@astro.livjm.ac.uk}

\author{S. P. S. Eyres}
\affil{Centre for Astrophysics, University of Central Lancashire, Preston PR1 2HE, UK}
\email{spseyres@uclan.ac.uk}

\author{T. J. O'Brien}
\affil{School of Physics and Astronomy, Jodrell Bank Observatory, University of Manchester,
Macclesfield SK 11 9DL, UK}
\email{tob@jb.man.ac.uk}

\begin{abstract}
The first low radio frequency ($<1.4$ GHz) detection of the outburst of
the recurrent nova RS Ophiuchi is presented in this letter. Radio emission
was detected at 0.61 GHz on day 20 with a flux density of $\sim 48$ 
mJy and at 0.325 GHz on day 38 with a flux density of $\sim 44$ mJy. This is in
contrast with the 1985 outburst when it was not detected at 0.327 GHz even on day 
66. The emission at low radio frequencies is clearly non-thermal and is 
well-explained by a synchrotron spectrum of index 
$\alpha \sim -0.8$ ($S \propto \nu^\alpha$)
suffering foreground absorption due to the pre-existing, ionized, warm, clumpy red giant 
wind.  The absence of low frequency 
radio emission in 1985 and the earlier turn-on of the radio flux in the current 
outburst are interpreted as being due to higher foreground absorption in 1985 
compared to that in 2006, suggesting that the overlying wind densities in 2006 are 
only $\sim 30\%$ of those in 1985.

\end{abstract}

\keywords{binaries: close - novae, cataclysmic variables - stars: individual
(RS Ophiuchi) - stars: supernovae - stars: radio continuum}

\section{Introduction}

The recurrent nova RS Ophiuchi was discovered to be in outburst on 2006
February 12.83 UT \citep{narumi},  reaching a magnitude of $V=4.5$. 
The previous recorded outbursts took place in 1898, 1933, 1958, 1967 and
1985 \citep{rosino, rosino1}, with possible outbursts in 
1907 \citep{schaeffer} and 1945 \citep{oppenheimer}.

The interacting binary system of RS Oph comprises an M giant and a hot
accreting white dwarf with an orbital period of $455.72\pm0.83$ days 
\citep{dobrzycka, anupama, fekel}.
The nova outbursts are powered by a thermonuclear runaway 
on the white dwarf surface following accretion of 
mass from the companion \citep{starrfield, kato}.
There is remarkable similarity between the optical light curves and
spectra from different outbursts (e.g. Rosino 1987) . The 1985 outburst of 
RS Oph was one of the best studied events, with the outburst being 
recorded from X-rays to radio wavelengths \citep{bode1}. 

The first detection of the radio emission from RS Oph was made 18 days after the 
1985 outburst on January 26, by Padin et al. \nocite{padin} (1985) who observed a rise in the 
flux density of the radio source over $\sim 20$ days and noted that the
data indicated high brightness temperatures which suggested a nonthermal origin. 
Monitoring of the 1985 outburst with the VLA at frequencies of 1.49, 4.85, 4.885, 
14.94 and 22.46 GHz \citep{hjellming} indicated that at least two radio 
components existed from roughly one to six months; one with a negative spectral 
index (where $S_\nu \propto \nu^\alpha$) between 1.4 and 5 GHz; and another with a 
positive spectral index above 5 GHz.
The positive index component completely dominated the decaying spectrum. Taylor et al. (1989)
suggested the emission to be thermal at the higher frequencies. An 
interesting feature to be noted in the RS Oph radio source is that while it was 
easily detected at all the higher frequencies ($\ge 1.4$ GHz) observed, it was not 
detected at 0.325 GHz \citep{spoelstra}.  The emission from the 1985 outburst was also extensively 
modelled \citep{bode,brien}.

The 2006 outburst of RS Oph has been subject to intense monitoring in 
all wavebands, almost immediately after discovery \citep{brien1, bode2, sokoloski, das}.
The radio emission was detected, at frequencies $\ge 1.4$ GHz as early as 4.7 days 
from outburst \citep{eyres1}. The radio source of RS Oph was resolved on day 
13.8 in the VLBI observations \citep{brien1}. The source, which was found 
to be an almost complete ring in the initial observations, soon evolved to a complex,
multi-component structure consisting of an equatorial ring and polar caps. The early
asymmetries in the radio source were explained by O'Brien et al. (2006) as being
due to foreground absorption. Based on the observed brightness temperature and the 
estimate of the density in the shell, O'Brien et al. (2006) concluded that the radio 
emission is dominated by a non-thermal synchrotron component. 

We present in this letter the low frequency radio observations of RS Oph
using the Giant Metrewave Radio Telescope (GMRT). We report the 
detection, for the first time, of RS Oph at frequencies below 1.4 GHz, and 
interpret the observed light curves.

\section{Observations}

Continuum observations with the GMRT \citep{swarup} were first made on 2006 
February 24 (day 12.62) at 1.28 GHz. 
Subsequently, monitoring of the nova was performed at 0.61, 0.325,   
0.24, and 0.15 GHz. Since the 0.61 and 0.24 GHz feeds at GMRT are concentric, we 
observed both the frequency bands simultaneously. The last observation at 0.61 GHz 
was on 30 Jan 2007, and on 24 Nov 2006 at 0.325 GHz. 3C286 was used as the flux 
calibrator. 

The data obtained in native `LTA' format were converted to standard FITS format and 
analysed using standard tasks in NRAO AIPS\footnote{AIPS is distributed by NRAO
which is a facility of the NSF operated under cooperative agreement by Associated
Universities, Inc. }.
Self calibration using wide-field imaging was used to 
improve the image quality and, on the average, two rounds of phase self calibration 
were found to give the best images.  Radio emission was detected at all the observed 
radio frequencies, except at 0.15 GHz observed on 2 June 2006 (day 110) 
to a $3\sigma$ limit of 27 mJy.  We detected radio emission from the source
at 0.325 and 0.61 GHz on all the observed days.  The typical beamsizes at these
two frequencies were $\sim 10''$ and $\sim 6''$ respectively and the 
source was not resolved in our observations.
The uncertainty in the flux scale is $\sim 15\%$ and the 
errors on the flux densities listed in Table \ref{tab1} reflect this 
uncertainty.

\section{The Radio Light Curves}
\begin{figure}
\includegraphics[width=17cm]{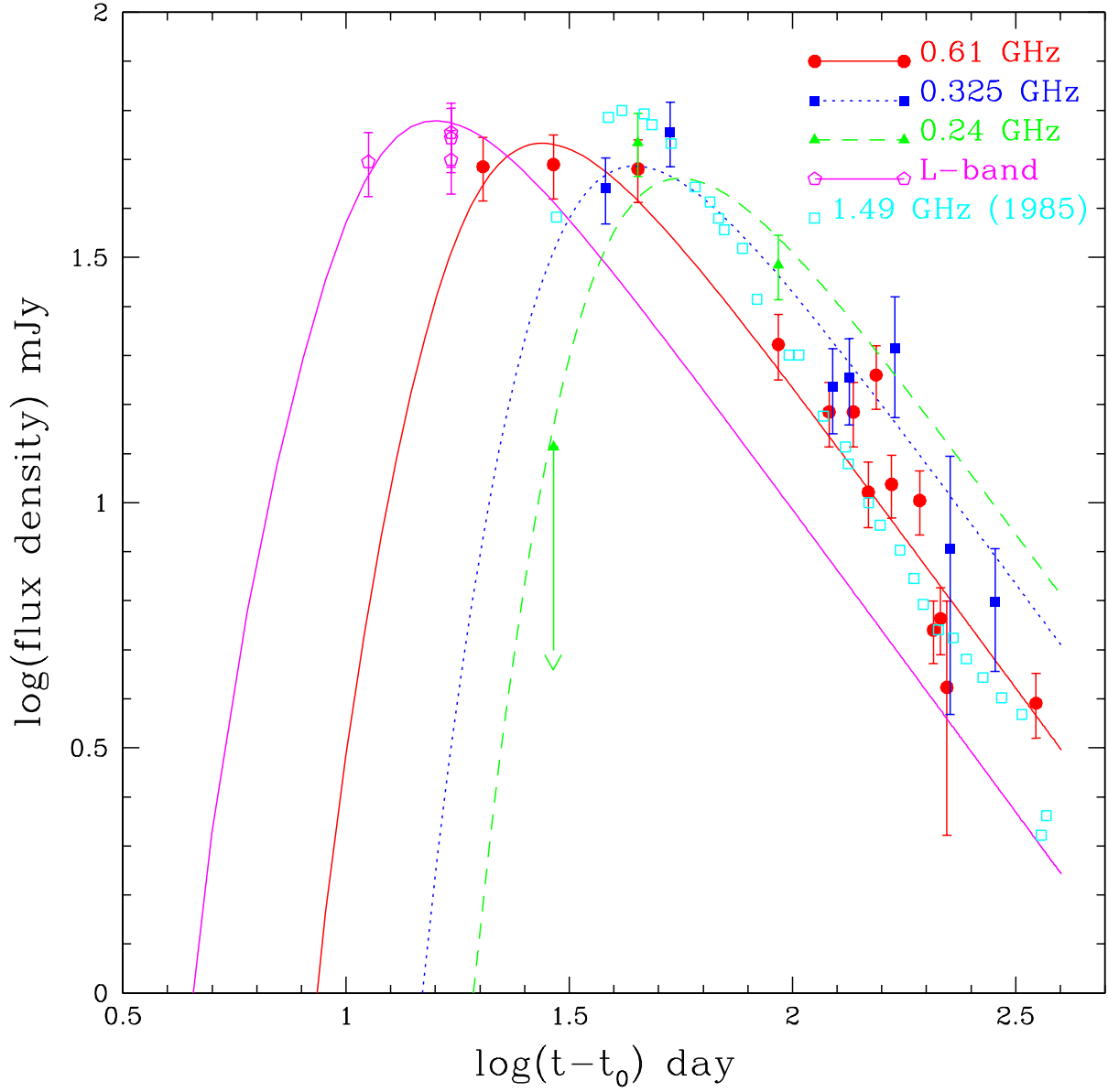}
\caption{Observed light curves of RS Oph (points) at L-band, 0.61 GHz,
0.325 and 0.24 GHz observed in the 2006 outburst. Continuous lines represent
the model light curves generated using parameters listed in Table \ref{tab2}.
Also plotted is the 1.49 GHz light curve from 1985 outburst \citep{hjellming}.}
\label{fig1}
\end{figure}

Fig. \ref{fig1} shows the flux density evolution of RS Oph at 0.24, 0.325 and 
0.61 GHz observed with the GMRT in the period $20-351$ days after the outburst. 
The first observations at 0.61 GHz, 0.325 GHz and 0.24 GHz were on March 5 (day 20), 
March 23 (day 38) and March 30 (day 47) respectively. Also plotted in Fig. \ref{fig1} 
are the L band observations using GMRT and the 1.49 GHz observations from the 1985 outburst. 

The light curves indicate a steep rise in the flux density, followed by 
a relatively flat maximum  and a subsequent  decay.  While the different frequencies 
seem to become visible and peak at different epochs with the lower frequencies turning on at 
later times which is clear from our upper 
limit at 0.24 GHz around day 25 and subsequent detection around day 45; the 
post-maximum decay at all the frequencies is fairly similar. 
The observed spectral index varies from $\alpha \sim -0.1$ around maximum
to $\alpha \sim -1.0$ around day 220.  These observations clearly show that
the radio emission at these frequencies is non-thermal.  Although the non-thermal
nature of the radio emission was inferred from the brightness temperatures and
comparison between radio and X-ray flux \citep{taylor} for the 1985 outburst, this is 
the first time that the non-thermal nature of the low frequency radio emission has been
clearly demonstrated by the negative spectral index. 

\section{Discussion}

The turn-on delay at longer wavelengths and the power-law decline after maximum
with index $\beta$ and a decreasing spectral index $\alpha$ observed in RS Ophiuchi
are very similar to the properties of radio supernovae, and imply the emission to
be nonthermal synchrotron. We model the observed light curves adopting the
models for radio supernovae \citep{weiler,weiler1} wherein the relativistic 
electrons and enhanced magnetic fields necessary for synchrotron emission are 
generated due to the shock interaction of the nova ejecta with the circumbinary 
red giant wind material that is ionized and heated by the nova explosion 
\citep{chevalier1,chevalier2}.  The radio 
emission rises rapidly as the shock progressively overtakes the wind material, 
causing a decrease in the line of sight absorption. Evidence for the presence 
of such an ionised, warm red giant wind is provided by the X-ray 
\citep{mason,bode2,sokoloski}, radio \citep{brien1} and optical 
(G.C. Anupama et al.\ 2007, in preparation, Bode et al. 2007) observations.

\subsection{Model light curves}

The following model, based on Weiler et al. (2002) \nocite{weiler1} is adopted. 

\begin{eqnarray}
S({\rm{mJy}})= K_1\,\left(\nu \over1{\rm GHz} \right)^\alpha\,\left(t-t_0\over20{\rm d}\right)^\beta\, e^{-\tau^{\rm{CSM}}_{\rm{homog}}}\,\left({1-e^{-\tau^{\rm{CSM}}_{\rm{clumps}}}}\over{\tau^{\rm{CSM}}_{\rm{clumps}}} \right) \rm{,}
\end{eqnarray}
where
\begin{eqnarray}
\tau^{\rm{CSM}}_{\rm{homog}}=K_2 \left(\nu\over{\rm 1GHz}\right)^{-2.1}\,\left(t-t_0\over20{\rm d}\right)^\delta\\
\tau^{\rm{CSM}}_{\rm{clumps}}=K_3 \left(\nu\over1{\rm GHz}\right)^{-2.1}\,\left(t-t_0\over20{\rm d}\right)^{\delta^\prime} .
\end{eqnarray}

In the above model, $t_0$ corresponds to 2006 Feb 12.83 and $t-t_0$ is time
since the outburst.  $K_1$ represents the flux density, $K_2$ the attenuation by
a homogeneous absorbing medium and $K_3$ the attenuation by a clumpy/filamentary
medium, at a frequency of 1 GHz, 20 days after the nova explosion.  
The optical depths, $\tau^{\rm{CSM}}_{\rm{homog}}$ and $\tau^{\rm{CSM}}_{\rm{clumps}}$
are due to the ionized circumstellar material (CSM) external to the emitting region.
$\alpha$ gives the spectral index and $\beta$ the rate of decline in the optically
thin phase. The optical depths in the homogeneous and clumpy/filamentary
CSM are described by $\delta$ and $\delta^\prime$ respectively. 
A simultaneous, non-linear chi-square fit using the Levenberg-Marquardt 
algorithm \citep{press} is made to the data at all the frequencies 
and the best fit parameters ($\chi_{\rm{red}}= 1.5$)
are determined as listed in Table \ref{tab2}.  
The flux density at 1.46 GHz on day 4.7 ($2.8\pm0.2$ mJy), reported by 
Eyres et al. \nocite{eyres1}(2006), is also used in the model fit. 
The resulting model light 
curves are plotted in Fig. \ref{fig1}. It is clear that the overall evolution of 
the radio emission at freq $< 1.4$ GHz is well described by the model.  The model 
predicts the delayed turn-on at different frequencies fairly well (ref.\ Table 
\ref{tab2}), implying the turn-on time is determined by the thermal 
optical depth of the CSM. However it underpredicts the peak fluxes at the
0.325 GHz and 0.24 GHz.  Using our model fits and 
Eqns (11) and (13) of Weiler et al. (2002),  we estimate a mass loss rate of
$1.6\times10^{-7} M_{\odot}~ {\rm yr}^{-1}$ for a red giant wind velocity of $20$ kms$^{-1}$
\citep{bode}. 

Excess emission, over the model fit, is observed at 0.61 and 
0.325 GHz beyond day $\sim 120$.
No such flux density variation is observed in two other bright sources in 
the field, implying the enhancement is intrinsic to RS Oph. 
It is interesting to note that S. P. S. Eyres et al.\ (2007, in preparation) 
record a slight enhancement in the flux density at
1.4 and 5 GHz on day 155. The enhancement that we record appears to be different from 
the short duration ($\la 24$~hrs) increase in the flux density that was observed at 5 GHz, 
41 days after the 1985 outburst and interpreted as being due to increased
thermal emission \citep{spoelstra}. 
To quantify the difference in the observed and modelled
light curves, we followed Ryder et al. (2004) \nocite{ryder} and estimated the 
deviations for all the frequencies.  
Although the residuals are noisy, excess emission by a factor of 1.8 to 1.4 is
observed between days 120 and 200. 
The deviations in the light curve may be interpreted as being due to 
modulations in the CSM and the deviations are expected to be pronounced for edge-on viewing 
of the binary \citep{ryder}.
The radio luminosity is related to the change in the CSM density 
\citep{chevalier1} and for the RS Oph system (with an inclination of
$30^{\circ}-40^{\circ}$; Dobrzycka \& Kenyon, 1994 \nocite{dobrzycka}), 
the observed enhancement can be explained if the CSM density is increased by 50\% to 25\%. 
It is interesting to note that although it has been proposed that the forward shock may
traverse the red giant wind in around 80 days (e.g Bode et al. 2006 \nocite{bode2}),
VLBI and HST observations \citep{brien1,bode3} are consistent with enhanced density
in the pre-outburst CSM in the plane of the central binary.  Shock breakout would therefore
be expected to occur much later from material in this plane and thus this could be where
the density enhancements inferred from our results reside. 
Alternately, the flux density enhancement could be related to the presence of a new short
lived ($\sim 100$ days) component in the radio emitting medium.
The magnetic field for such a component, estimated assuming equipartition and the break frequency
to be 0.325 GHz would be $\sim 3.5$~G. This is much higher than the equipartition
magnetic fields in the remnant which were estimated in the 1985 outburst
to be of the order of $0.01-0.05$~G
\citep{bode,taylor}.  Spoelstra et al. (1987) have noted that a uniform
magnetic field can be enhanced by a factor of 200 if the cooled remnant material
is compressed in volume by a factor of 100 as the nova remnant expands. When the
shocked medium becomes well cooled, Rayleigh-Taylor instabilities can cause such
clumps to form.

\subsection{Comparison with 1985 radio light curves}

In Fig. \ref{fig1}, the light curve for 1.49 GHz from the 1985 outburst 
\citep{hjellming} is plotted alongwith with the present low frequency light curves. The sharp rise to 
peak at 0.325 and 0.24 GHz is very similar to the sharp rise to peak observed in 
1985. The peak at 0.61 GHz is broader than at other frequencies. Although 
differences are seen around the maximum phase, it is interesting to note the very 
similar evolution during both outbursts in the post-maximum phase, 
particularly after day $\sim 60$. 

We note two major differences in the radio emission during the 1985 and 2006 outburst:

(i) The detection of emission at frequencies $<1.4$ GHz. RS Oph was
not detected at 0.325 GHz on day 48 down to a $1\sigma$ limit of 
5 mJy during the 1985 outburst \citep{spoelstra}, while it was 
detected close to maximum on day 20 at 0.61 GHz and on day 38 at 
0.325 GHz during the current outburst. 

(ii) The detection of radio emission as early as day 4.7 \citep{eyres1}
compared to an implied turn-on at day 14 in 1985 (e.g.\ Padin et al.\ 1985). 

Both these observations suggest that a steep-spectrum synchrotron source
became visible at the lower frequencies during the current outburst due to reduced 
foreground absorption compared to the 1985 outburst.  
Using the $3\sigma$ limit of 12~mJy on day 56 for the 0.325 GHz emission in 1985 \citep{spoelstra}, 
the observed value of 57 mJy on day 53 in 2006 and assuming that only foreground uniform
absorption is responsible for the difference, we find a relation for 
the optical depths at 0.325 GHz in the two epochs 
$\tau(1985) \ge \tau(2006) + 1.6$ from simple radiative transfer arguments. 
The light curve model indicates that 
that $\tau(2006)\sim 0.2$ on day 53 which implies $\tau(1985) \sim 1.8$. 
Since $\tau \propto n_e^2~L$,
this indicates that the electron density
of the foreground gas in 2006 is about 30\% of that in 1985 for the same linear depth $L$.
Since the two outbursts are otherwise fairly similar \citep{brien1}, 
one of the reasons for this difference could be 
the variation in the clumpiness of the foreground gas.  Observations 
of RS Oph at quiescence do indicate the presence 
of non-uniform column density in the red giant wind envelope, that is uncorrelated 
with the binary geometry \citep{anupama}.
The non-detection at 0.325 GHz of the 1985 outburst had been attributed 
to absorption by the thermal gas mixed with the synchrotron emitting gas 
and low frequency cut-off in the electron spectrum \citep{spoelstra}.  However 
both the physical processes appear to be negligible in the 2006 outburst.  
 
\section{Summary}
GMRT observations of RS Oph during the 2006 outburst have shown early emission 
at frequencies below 1.4 GHz for the first time. The emission is clearly 
non-thermal, synchrotron emission and the evolution appears to be similar to 
that observed for radio supernovae. 
The light curves at these low frequencies are well explained 
by decreasing free-free absorption by the foreground CSM of the 
synchrotron emission from the remnant of the nova outburst.
Model light curves with a spectral index of 
$\alpha =-0.8$, decay power law of index $\beta=-1.2$ and 
optical depth due to a homogeneous, clumpy absorbing medium well fit the observed 
light curves. Clumpy medium seems to dominate the absorption at early times. 
The light curves indicate the appearance of emission components
consistent with the VLBA, EVN and MERLIN images \citep{brien1} at higher 
frequencies.  Modelling this data with data at
higher frequencies will enhance our understanding of this unique recurrent nova system.

The similarity of late time evolution of the nova during the 1985
and 2006 outbursts suggests that the early difference is primarily due
to the free-free absorption in the ionized CSM. This
makes a case for early multifrequency observations with good
temporal sampling during future outbursts.

\acknowledgements
{We thank the anonymous referee for insightful comments on the light curve modelling which have
enhanced and improved the results of this paper.  We thank Prof. R. Nityananda for providing
TOO time on the GMRT during the early stages of the nova outburst and NGK thanks him
for discussions.  We thank the staff of the GMRT that made these observations possible.
GMRT is run by the NCRA, a centre of the TIFR.  This research has made use of 
NASA's Astrophysics Data System.
}

\clearpage

\begin{deluxetable}{lllll}
\tablewidth{0pt}
\tablecaption{Observed flux densities for RS Ophiuchi.   
The first column indicates
the day after the outburst, 2006 Feb 12.83 UT (JD\,2453779.33). }
\tablehead{
\colhead{Day}&\colhead{$\nu$} & \colhead{S} &\multicolumn{2}{c}{Phase cal}\\
\multicolumn{1}{l}{} &\colhead{GHz}  &\colhead{ mJy}  & \colhead{Name} & \colhead{S (Jy)} 
}
\startdata
17.19 & 1.39 & $56.8 \pm 8.5$ & 1743-038 & $3.5\pm 0.5$ \\
11.22 & 1.28 & $49.5 \pm 7.4$ & 1743-038 & $2.7\pm 0.4$ \\
17.19 & 1.28 & $50.0 \pm 7.5$ & 1743-038 & $2.7\pm 0.4$ \\
17.19 & 1.06 & $55.4 \pm 8.3$ & 1743-038 & $2.3\pm 0.4$ \\
\\
20.26 & 0.61 & $48.4 \pm 7.2$ & 1743-038 & $1.1 \pm 0.2$ \\
29.15 & 0.61 & $48.9 \pm 7.3$ & 1833-210 & $9.8 \pm 1.5$ \\
45.07 & 0.61 & $47.9 \pm 7.2$ & 1822-096 & $7.7 \pm 1.2$ \\
92.99 & 0.61 & $21.0 \pm 3.2$ & 1822-096 & $8.1 \pm 1.2$  \\
120.86 & 0.61 & $15.3 \pm 2.3$ & 1822-096 & $8.4 \pm 1.3$  \\
136.92 & 0.61 & $15.3 \pm 2.3$ & 1822-096 & $8.5 \pm 1.3$ \\
147.88 & 0.61 & $10.5 \pm 1.6$ & 1822-096 & $8.8 \pm 1.3$ \\
153.94 & 0.61 & $18.2 \pm 2.7$ & 1822-096 & $9.0  \pm 1.4$ \\
166.77 & 0.61 & $10.9 \pm 1.6$ & 1822-096 & $10.0 \pm 1.5$\\
192.84 & 0.61 & $10.1 \pm 1.5$ & 1822-096 & $9.9 \pm 1.5$ \\
206.89 & 0.61 & $5.5 \pm 0.8$ & 1822-096 & $8.2 \pm 1.2$\\
214.66 & 0.61 & $5.8 \pm 0.9$ & 1822-096 & $8.3 \pm 1.2$\\
221.86 & 0.61 & $4.2 \pm 2.1$ & 1822-096 & $9.2 \pm 1.4$\\
351.40 & 0.61 & $3.9 \pm 0.6$ & 1822-096 & $7.8 \pm 1.2$\\
\\
38.15 & 0.325 & $43.7 \pm 6.6$ & 1822-096 & $10.4 \pm 1.6$\\
53.24 & 0.325 & $57.0 \pm 8.6$ & 1822-096 & $11.8 \pm 1.8$\\
123.12 & 0.325 & $17.2 \pm 3.4$ & 1822-096 & $9.8 \pm 1.5$\\
134.07 & 0.325 & $18.0 \pm 3.6$ & 1822-096 & $10.7 \pm 1.6$\\
169.75 & 0.325 & $20.6 \pm 5.7$ & 1822-096 & $10.1 \pm 1.5$\\
225.61 & 0.325 & $8.1 \pm 4.4$ & 1822-096 & $9.5 \pm  1.4$\\
284.43 & 0.325 & $6.3 \pm 1.8$ & 1822-096 & $9.7 \pm 1.5$\\
\\
29.15 & 0.24 & $<13.0$    & &\\
45.07 & 0.24 & $54.2 \pm 8.0$ & 1822-096 & $14.0 \pm 2.1$\\
92.99 & 0.24 & $30.5 \pm 4.6$ & 1822-096 & $15.0 \pm 2.3$\\
\\
110 & 0.15 & $<9.0$   &  &\\
\enddata
\label{tab1}
\end{deluxetable}

\clearpage
\begin{deluxetable}{lc}
\tablewidth{0pt}
\tablecaption{Best fit parameters from least square fits to the low frequency 
radio emission from RS Ophiuchi}
\tablehead{
\colhead{Parameter} & \colhead{Value} 
}
\startdata

$K_1$ & $86.4$\\
$\alpha$ & $-0.78$\\
$\beta$ & $-1.24$\\
$K_2$ & $0.144$\\
$\delta$ & $ -2.29$\\
$K_3$ & $0.53$\\
$\delta^\prime$ & $-3.14$\\
\hline
{$\nu$} (GHz) & Onset (days)  \\
\hline
$1.28$ & 3 \\
$0.61$ & 6 \\
$0.325$ & 10 \\
$0.24$ & 14 \\
\enddata
\label{tab2}
\end{deluxetable}

\clearpage

\begin{thebibliography}{}
\bibitem[Anupama \& Miko{\l}ajewska 1999]{anupama} Anupama, G. C., \& Miko{\l}ajewska, J. 1999, A\&A, 344, 177
\bibitem[Bode \& Kahn 1985]{bode} Bode, M. F., \& Kahn, F. D. 1985, MNRAS, 217, 205
\bibitem[Bode 1987]{bode1} Bode, M. F.,  1987, Proceedings of the meeting on
RS Ophiuchi (1985) and the Recurrent Nova Phenomenon, Utrecht: VNU Science Press,  
(Ed: Bode, M.F)
\bibitem[Bode et al 2006]{bode2} Bode, M. F., et al. 2006, ApJ, 652, 629
\bibitem[Bode et al 2007]{bode3} Bode, M. F., Harman, D. J.,  O'Brien, T. J., Bond, H. E., 
Starrfield, S., Darnley, M. J., Evans, A., Eyres, S. P. S.  2007, ApJL, 665, L63
\bibitem[Chevalier 1982a]{chevalier1} Chevalier, R. A., 1982, ApJ, 259, 302
\bibitem[Chevalier 1982b]{chevalier2} Chevalier, R. A., 1982, ApJL, 259, L85
\bibitem[Das et al. 2006]{das} Das, R., Bannerjee, D. P. K., \& Ashok, N. M. 2006, ApJ, 653, L141
\bibitem[Dobrzycka \& Kenyon 1994]{dobrzycka} Dobrzycka, D., \& Kenyon, S. J. 1994, AJ, 108, 2259
\bibitem[Eyres et al. 2006]{eyres1} Eyres, S. P. S., O'Brien, T. J., Muxlow, T. W. B., Bode, M. F., Evans, A. 2006, IAUC 8678 
\bibitem[Fekel et al. 2000]{fekel} Fekel F. C., Joyce, R. R., Hinkle, K. H., Skrutskie, M. F. 2000, AJ, 119, 1375
\bibitem[Narumi et al. 2006]{narumi} Narumi, H.,  Hirosawa, K., Kanai, K., Renz, W.,
Pereira, A., Nakano, S., Nakamura, Y., Pojmanski, G., 2006,  IAUC 8671
\bibitem[Hjellming et al. 1986]{hjellming} Hjellming, R. M., van Gorkom, J. H.,
Seaquist, E. R., Taylor, A. R., Padin, S., \& Davis, R. J., et al. 1986, ApJ, 305, L71
\bibitem[Kato 1990]{kato} Kato, M. 1990, ApJ, 355, 277
\bibitem[Mason et al. 1987]{mason} Mason, K. O., Cordova, F. A., Bode, M. F., \& Barr, P.  1987, 
in Proceedings of the meeting on RS Ophiuchi (1985) and the Recurrent Nova Phenomenon, 
Ed: by M.F. Bode. Utrecht: VNU Science Press,  p167
\bibitem[O' Brien et al. 1992]{brien} O'Brien, T. J.,  Bode, M. F., \& Kahn, F. D., 1992, MNRAS, 255, 683
\bibitem[O' Brien et al. 2006]{brien1} O'Brien, T. J., et al. 2006, Nature, 442, 279
\bibitem[Padin et al. 1985]{padin} Padin, S., Davis, R. J., \& Bode, M. F. 1985, Nature, 315, 306
\bibitem[Press et al. 2002]{press} Press, W. H., Teukolsky, S. A., Vetterling, W. T., \& Flannery, B. P.,
2002, The Numerical Recipes, (2ed: Cambridge University Press)  
\bibitem[Rosino 1987]{rosino} Rosino, L. 1987, in  Proceedings of the meeting 
on RS Ophiuchi (1985) and the Recurrent Nova Phenomenon, Ed: M.F. Bode. 
Utrecht: VNU Science Press,  p1
\bibitem[Rosino \& Iijima 1987]{rosino1} Rosino, L. \& Iijima, T. 1987 in  Proceedings of the 
meeting on RS Ophiuchi (1985) and the Recurrent Nova Phenomenon, 
Ed: M.F. Bode. Utrecht: VNU Science Press,  p27
\bibitem[Ryder 2004]{ryder} Ryder, S. D., Sadler, E. M., Subrahmanyan, R., 
Weiler, K. W., Panagia, N., Stockdale, C. 2004, MNRAS, 349, 1093 
\bibitem[Schaeffer 2004]{schaeffer} Schaeffer, B. 2004, IAUC 8396
\bibitem[Oppenheimer \&  Mattei 1993]{oppenheimer} Oppenheimer, B. D. \& Mattei, J. A. 
1993, JAAVSO, 22, 105
\bibitem[Spoelstra et al. 1987]{spoelstra} Spoelstra, T. A. T., Taylor, A. R., Pooley, G. G.,
Evans, A., \& Albinson, J. S. 1987, MNRAS, 224, 791
\bibitem[Sokoloski et al. 2006]{sokoloski} Sokoloski, J. L., Luna, G. J. M., Mukai, K., \&
Kenyon, S. J. 2006, Nature, 442, 276
\bibitem[Starrfield et al. 1985]{starrfield} Starrfield, S., Sparks, W. M., Truran, J. W.  1985,
ApJ, 291, 136 
Ed: Bode, M. F. \& Evans, A., Wiley, Chichester, p39
\bibitem[Swarup et al. 1991]{swarup} Swarup, G., et al. 1991, Current Science, 60, 95
\bibitem[Taylor et al. 1989]{taylor} Taylor, A. R.,  Davis, R. J., Porcas, R. W., \&  Bode, M. F.  1989,
MNRAS, 237, 81
\bibitem[Weiler et al. 1986]{weiler} Weiler, K. W., Sramek, R. A., Panagia, N., van der Hulst, J. M.,
\& Salvati, M. 1986, ApJ, 301, 790
\bibitem[Weiler et al. 2002]{weiler1} Weiler, K. W., Panagia, N., Montes, M. J.,\& Sramek, R. A. 2002, ARA\&A, 40, 387
\end{thebibliography}
\end{document}